\documentclass[3p,times]{elsarticle}

\usepackage{ecrc}

\volume{00}

\firstpage{1}

\journalname{Nuclear Physics A}

\runauth{Anne M. Sickles}

\jid{nupha}

\jnltitlelogo{Nuclear Physics A}

\usepackage{graphicx}
\usepackage{amsmath,amssymb}

\newsavebox{\bigleftbox}

\newcommand{\pA}{\mbox{$p$$+$A} }
\newcommand{\dAu}{\mbox{$d$$+$Au} }
\newcommand{\pT}{\mbox{$p_T$} }
\newcommand{\pPb}{\mbox{$p$$+$Pb} }
\newcommand{\pp}{\mbox{$p$$+$$p$} }
\newcommand{\dA}{\mbox{$d$$+$Au} }
\newcommand{\pAu}{\mbox{$p$$+$Au} }
\newcommand{\PbPb}{\mbox{Pb$+$Pb} }

\begin{document}

\begin{frontmatter}



\title{Experimental Results on p(d)+A Collisions at RHIC and the LHC}

\author{Anne M. Sickles}
\address{Department of Physics, University of Illinois at Urbana-Champaign, Urbana, Illinois\\
Physics Department, Brookhaven National Laboratory, Upton, New York}




\begin{abstract}
Recent experimental results at both the LHC and RHIC show evidence
for hydrodynamic behavior in proton-nucleus and deuteron-nucleus collisions ($p$$+$A).
This unexpected finding has prompted new measurements in \pA collisions in
order to understand whether matter with similar properties is created in A+A and \pA collisions
or whether another explanation is needed.  In this proceedings, we will discuss the
new experimental data and its interpretation within the context of heavy-ion collisions.  
\end{abstract}

\begin{keyword}

\end{keyword}

\end{frontmatter}



\section{Introduction}
Collisions of protons or deuterons with large nuclei (hereafter collectively refered to as \pA collisions) 
were studied as control measurements for heavy-ion collisions.  Results since Quark Matter 2012 have
cast doubt on this paradigm.  The possible creation of hot nuclear matter
has been suggested by measurements of long range azimuthal correlations in \pPb collisions at 
the LHC and \dAu collisions at RHIC.  
These observations have motivated  an abundance of $v_N$ and
spectra measurements in small collision systems to determine to what extent a hydrodynamic description 
is applicable in \pA.
Additionally, significant modifications to jets and high-\pT hadrons have been observed at the LHC which remain
unexplained.
 In this proceedings we discuss a selection of the interesting experimental results related to \pA 
physics and highlight some of the open questions and new measurements that are needed.  

The results discussed here are divided into three sections: high-\pT jets and hadrons, identified particles,
and correlations.

\section{High-\pT Jets and Hadrons}
One of the original motivations for \pA physics at RHIC and the LHC is to quantify nuclear modifications 
to the production of hard observables as a baseline for heavy-ion measurements.
First results measurements of high-\pT single hadron production in both \dAu~\cite{Adams:2003im,Adler:2003ii}
 and \pPb~\cite{ALICE:2012mj} were consistent with binary scaled \pp to ~\pT$\approx$~10~GeV/$c$ at RHIC and
to $\approx$~20~GeV/$c$ at the LHC.  
New results presented at this conference show that reconstructed jet spectra are approximately
 consistent with binary scaling 
to very high \pT~\cite{Aiola:qm14,Appelt:qm14,Perepelitsa:qm14}.  
There is evidence for a slight excess of the nuclear modification factor, $R_{pPb}$, 
($<$~20\% at midrapidity) which is consistent with expectations from 
nuclear modifications to the parton distribution functions~\cite{Eskola:2009uj}.

\begin{figure}


\includegraphics[width=0.35\textwidth]{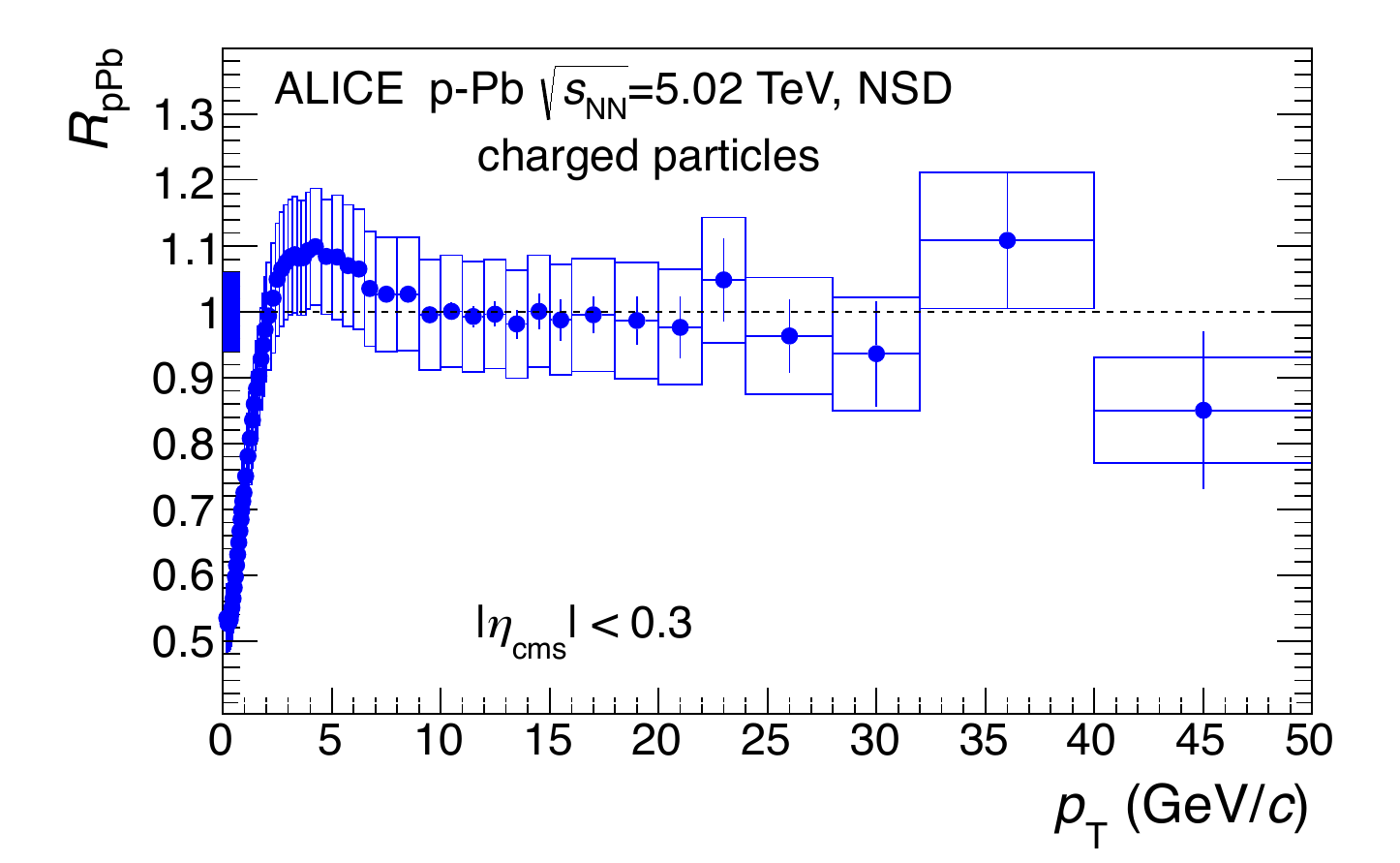}
\includegraphics[width=0.3\textwidth]{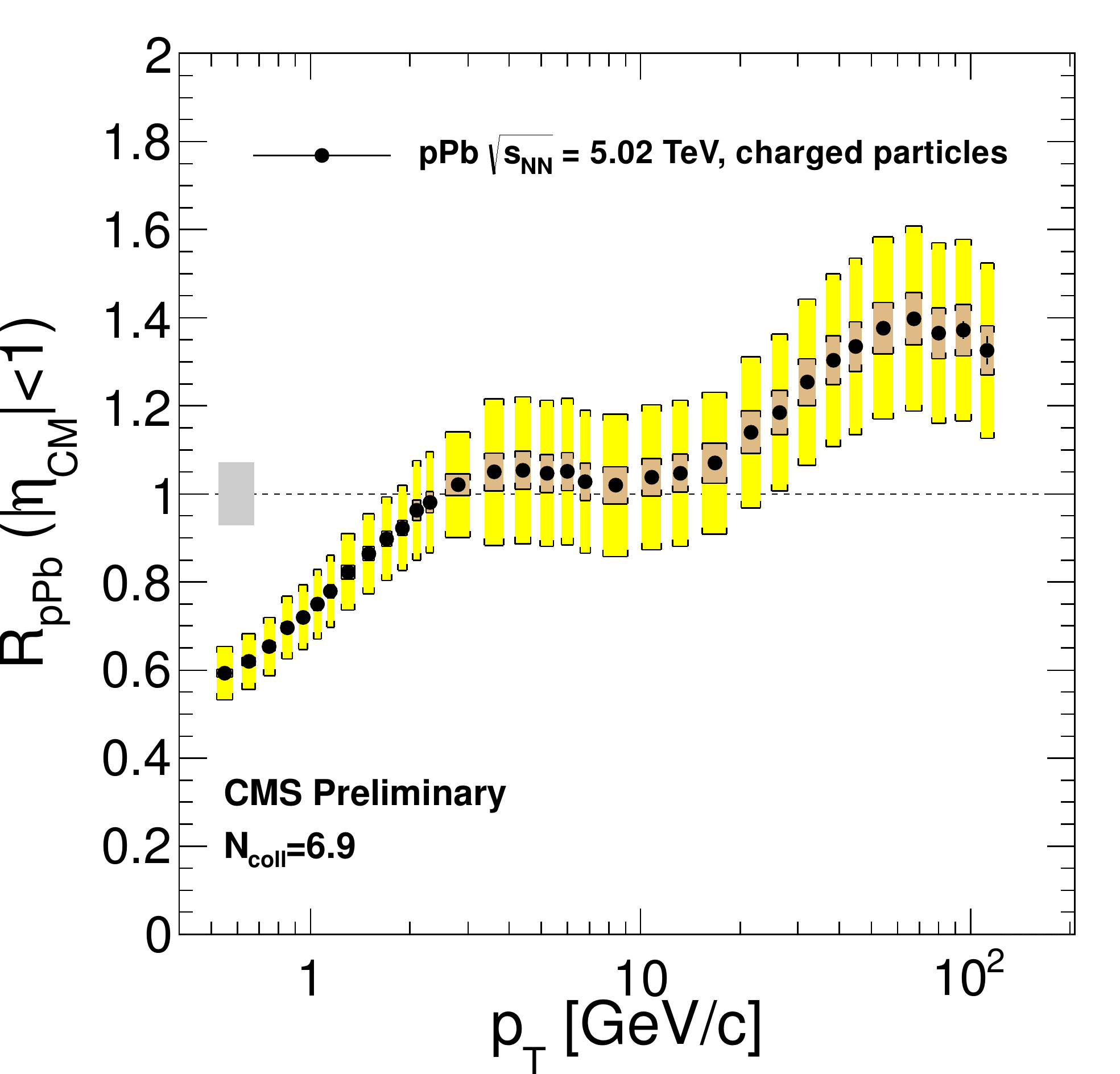}
\includegraphics[width=0.3\textwidth]{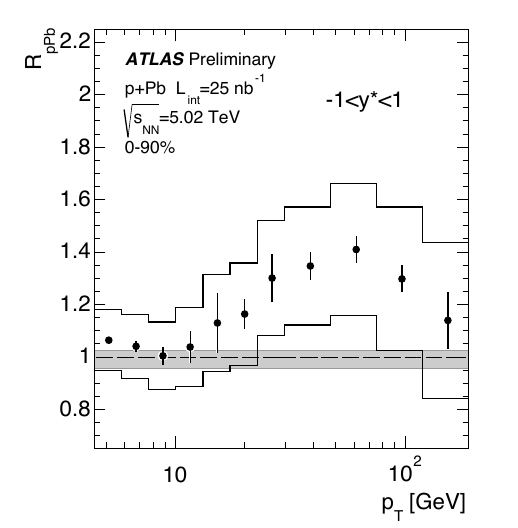}
\caption{Charged hadron $R_{pPb}$ at $\sqrt{s_{NN}}$~=~5.02~TeV from ALICE~\cite{Aiola:qm14}, CMS~\cite{Appelt:qm14}
and ATLAS~\cite{Balek:qm14} (from left to right).}
\label{fig:rppb}
\end{figure}

Surprisingly, single charged hadrons with $p_T>$~30~GeV/$c$ show a large and significant excess 
compared to binary scaling of \pp data
(up to approximately a 40\% increase) in recent measurements from the 
CMS~\cite{Appelt:qm14} and ATLAS~\cite{Balek:qm14} collaborations.  Results from the ALICE
experiment~\cite{Aiola:qm14} do not show this excess however the \pT reach is smaller and the statistical and
systematic uncertainties are large in the region where the excess is observed in CMS and ATLAS.
Results from all three collaborations are shown in Fig.~\ref{fig:rppb}.
The excess is larger than is expected from the jet data and larger and with a different \pT 
dependence than expected from the EPS09~\cite{Eskola:2009uj} nuclear parton distribution functions~\cite{Appelt:qm14}

It is notable that
there is no \pp data at 5.02~TeV, 
so there are substantial uncertainties in the \pp reference, which has to be interpolated from higher and
lower energy \pp collisions.
An analysis by CMS has found that much of the difference
between the ALICE and CMS results is in the \pp reference, not the \pPb measurement~\cite{Appelt:qm14}.  \pp running at 5~TeV is
of interest to establishing the excess observed by CMS and ATLAS and is central to reducing the uncertainties which
will enable more precise characterization of the charged hadron scaling.  In the short term, understanding the 
origin of the reference differences and direct comparison of the \pPb spectra are of interest.
The results from ATLAS and CMS
could suggest some modification to jet fragmentation in \pPb.  
It could also suggest that there is a larger fraction of quark jets than would be expected based on \pp collisions.
That might create an excess because of the softer fragmentation of gluon jets than quark jets, but
it has not been demonstrated that this would be  able to account for 
the size and \pT dependence of the charged hadron excess.

Centrality in \pA collisions has been measured by the forward 
energy in the nucleus going direction~\cite{Adare:2013nff,Perepelitsa:qm14,Toia:qm14}.
The centrality dependence of very high-\pT jet production in \pPb has been measured by ATLAS~\cite{Perepelitsa:qm14}
and they find an excess of jets in peripheral collisions and a suppression of jets in central collisions 
relative to binary scaling.  These
effects cancel to produce the $R_{pPb}$ for minimum bias collisions that is nearly unity.  This is 
similar to what was previously seen by PHENIX~\cite{Perepelitsa:2013jua}.  ATLAS has 
measured over
a wide range in jet pseudorapidity and finds that for a large span in $\eta$ the binary collision
scaled central to periphal jet ratio, $R_{CP}$ follows a constant trend as a function of 
the total jet momentum (not \pT).

\begin{figure}
\centering
\includegraphics[width=0.3\textwidth]{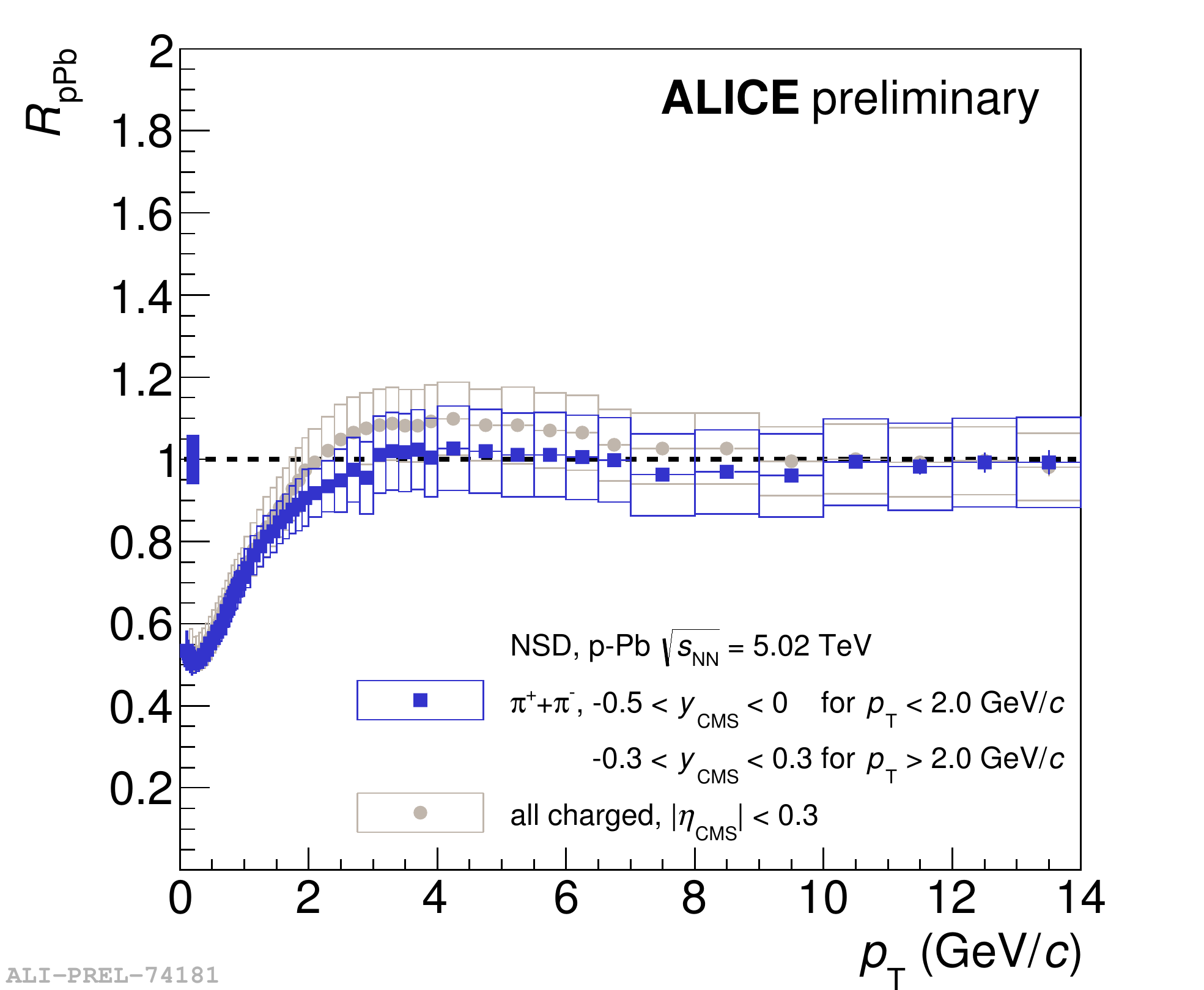}
\includegraphics[width=0.3\textwidth]{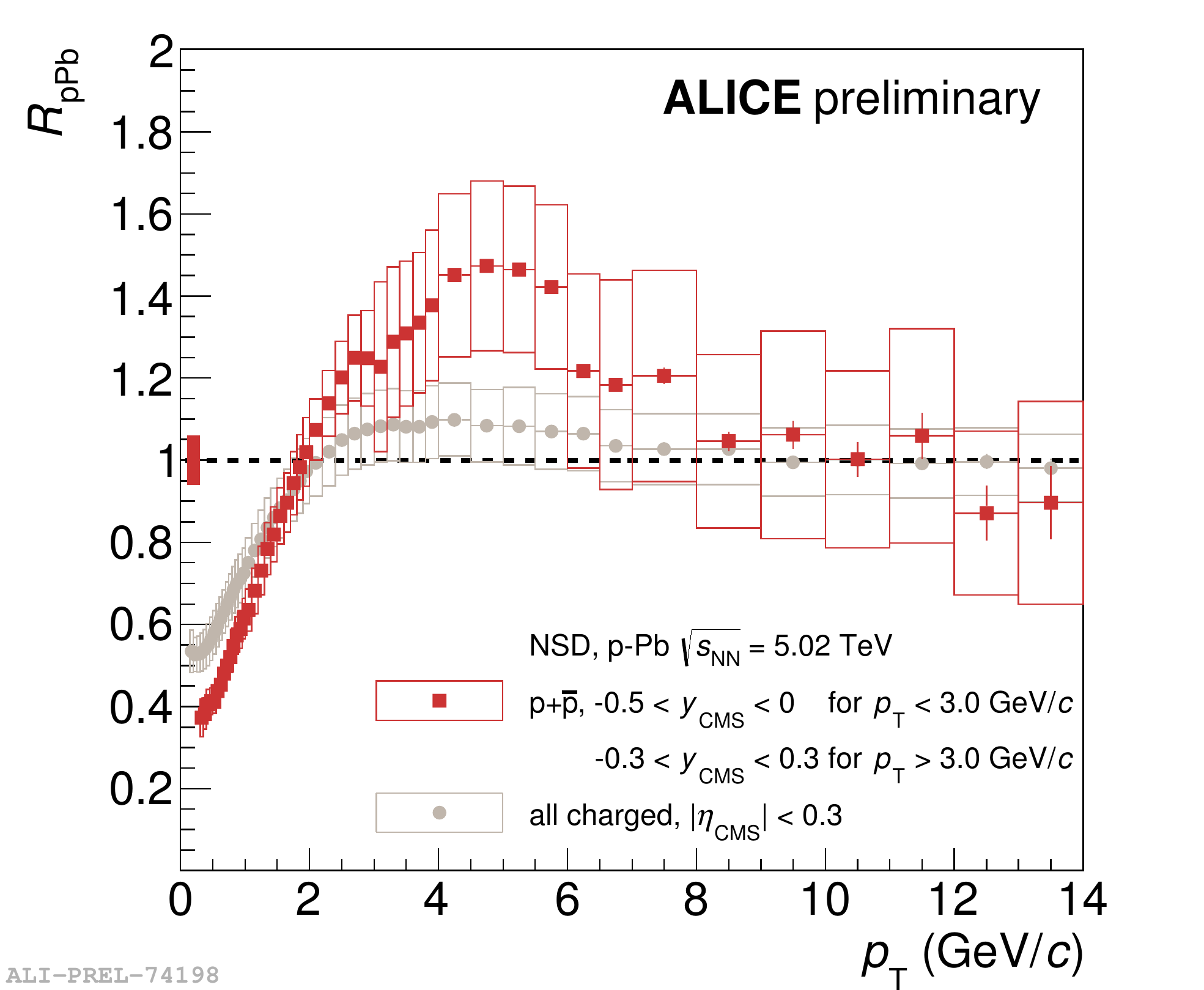}
\includegraphics[width=0.3\textwidth]{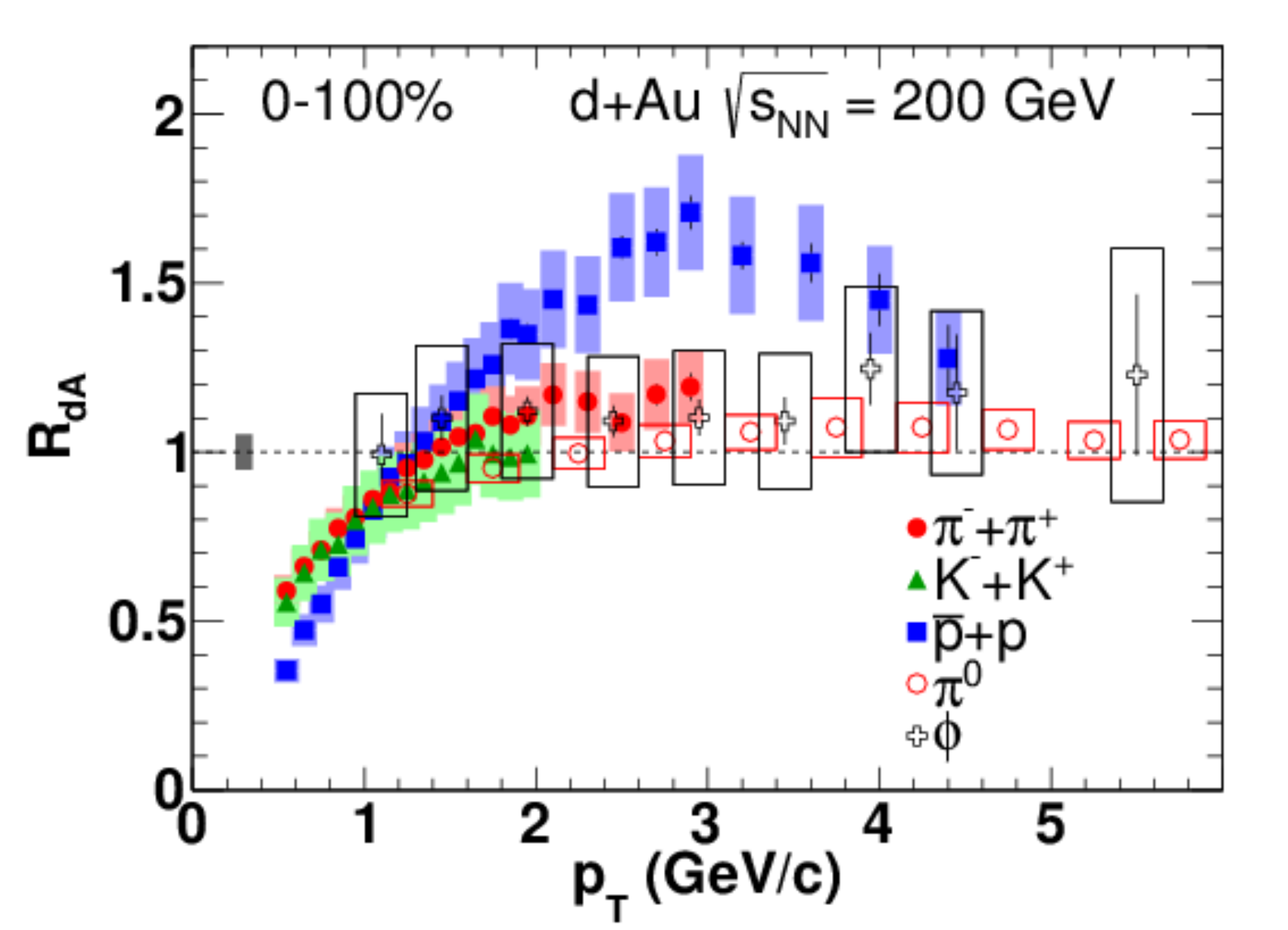}
\caption{$R_{pPb}$ for charged pions (left) and (anti-)protons (middle) for minimum
bias \pPb collisions~\cite{Knichel:qm14}.  Pions are consistent with unity for $p_T>$~4~GeV/$c$
and the (anti-)protons are enhanced by approximately 20--40\% for 3~$<p_T<$~6~GeV/$c$.
Similar measurements in \dAu collisions at $\sqrt{s_{NN}}$~=~200~GeV are shown in the right
panel~\cite{Adare:2013esx}.  Pions are consistent with unity and (anti-)protons are enhanced by approximately 60\%.} 
\label{fig:pidrpA}
\end{figure}
\section{Identified Particle Measurements in \pA}
\label{sec:spectra}

The particle spectra at lower \pT in \dA collisions have shown an increased yield of
protons and anti-protons compared to pions for 1.5~$<p_{T}<$~3~GeV/$c$~\cite{Adams:2003qm,Adler:2006xd}.  New
data from PHENIX~\cite{Adare:2013esx} and ALICE~\cite{Knichel:qm14} extend those measurements.
The nuclear modification factors for pions and protons at both PHENIX and ALICE are shown
in Fig.~\ref{fig:pidrpA}.
In order to further understand the origin of the  modified particle ratios, ALICE measured 
the $(\Lambda + \bar{\Lambda})/2K^0_S$ ratio of both inclusive particles and those 
associated with reconstructed jets in \pPb collisions~\cite{Zhang:qm14}.  The
inclusive $(\Lambda + \bar{\Lambda})/2K^0_S$ ratio is three times larger than that which
is measured in reconstructed jets.  This suggests that the species dependence of $R_{pA}$
is not associated with jets.

Since the particle species dependent modifications seem to be associated with non-jet 
particle production it is interesting to determine if hydrodynamics can account for the 
patterns in the data. The ALICE collaboration
has preformed blast-wave~\cite{Kolb:2000fha,Schnedermann:1993ws} fits to the \pPb spectra and find the different particle
species are well described by a single set of blast-wave parameters and that the extracted 
velocities increase as function of the multiplicity of produced particles in the collision~\cite{Abelev:2013haa}.

\begin{figure}
\centering
\includegraphics[width=0.3\textwidth]{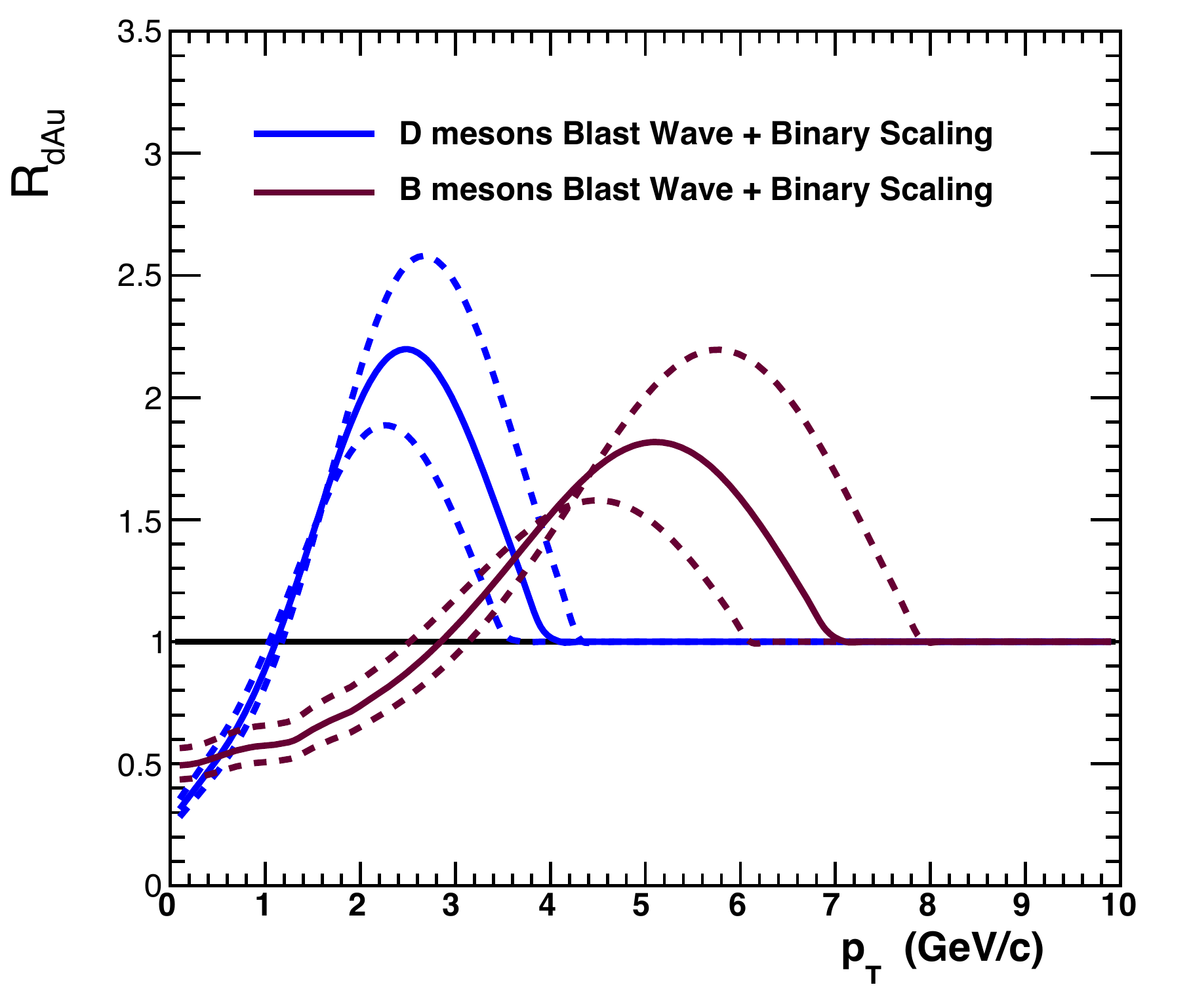}
\includegraphics[width=0.27\textwidth]{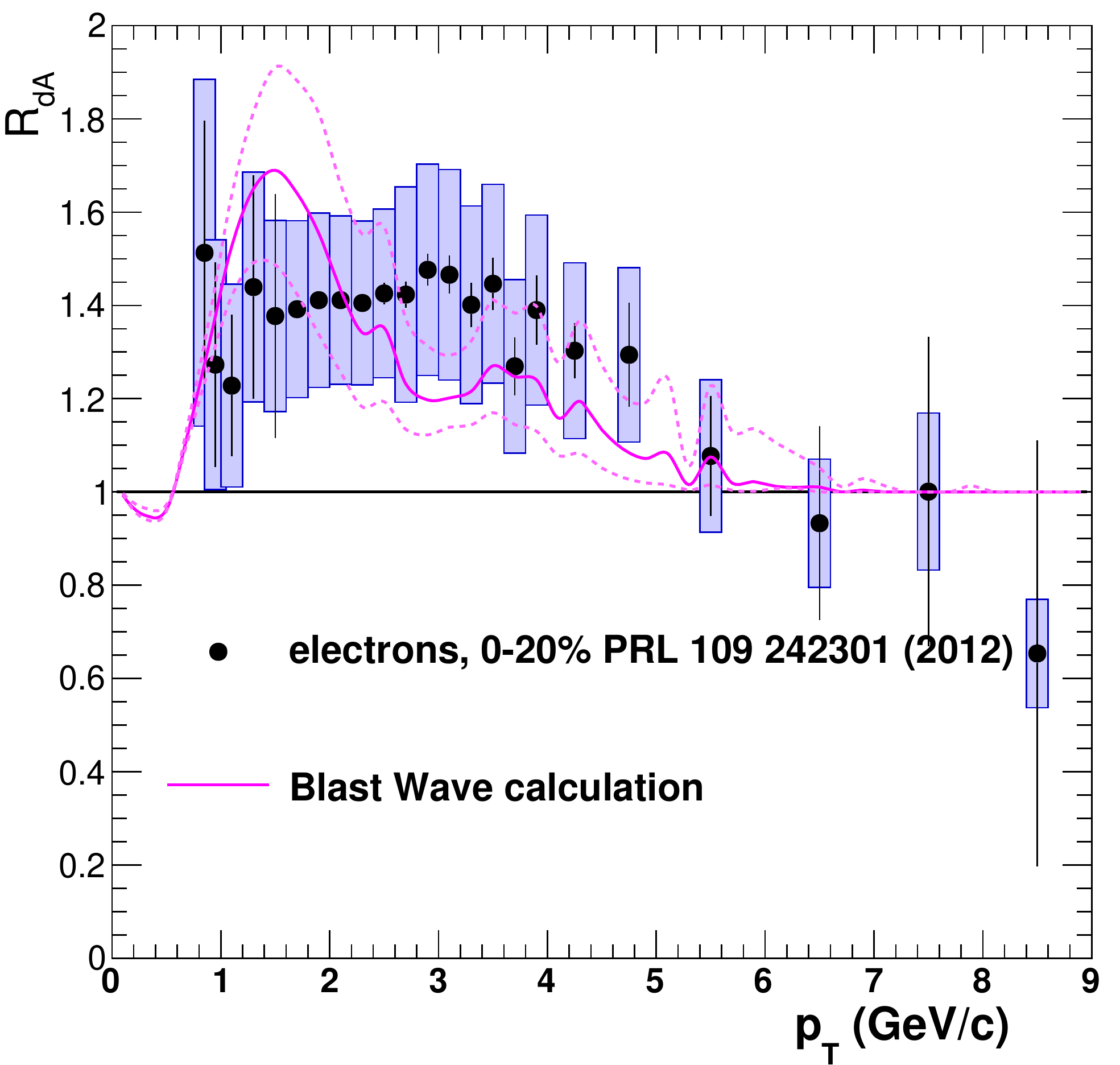}
\includegraphics[width=0.29\textwidth]{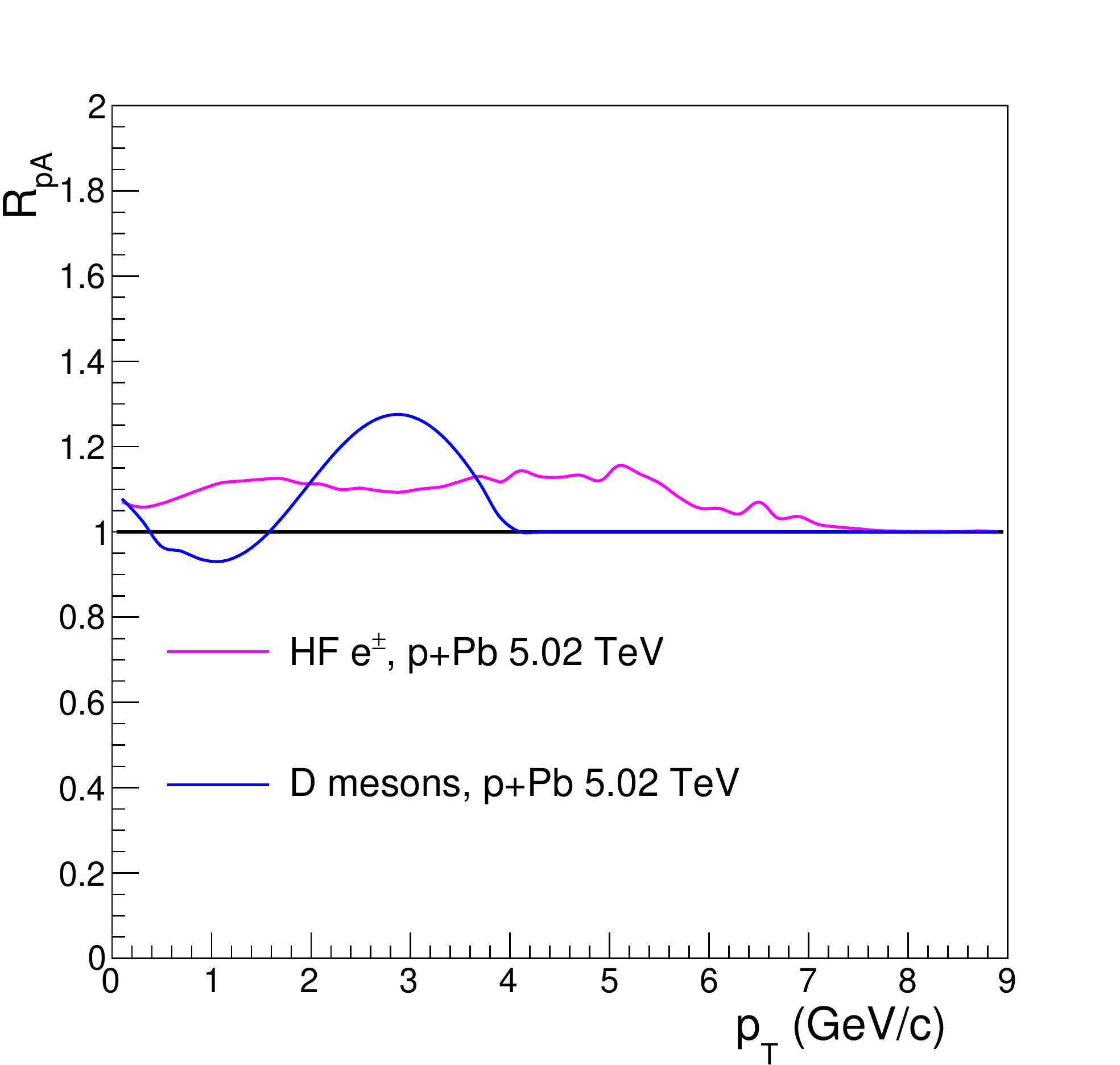}
\caption{(left) Blast-wave expectations for $D$ and $B$ meson nuclear modification factors at RHIC. (center) The resulting 
nuclear modification factor for 
electrons from the decay of $D$ and $B$ mesons (curves) and heavy meson decay electron
measurement in central \dAu collisions (points)~\cite{Adare:2012yxa}. (right) The same quantity for
$D$ mesons and  decay electrons at the LHC.  Figures are from 
Ref.~\cite{Sickles:2013yna}.}
\label{fig:bwrda}
\end{figure}
In light of the description of light hadrons at the LHC, it is  natural to ask whether this
descriptions extends to $\sqrt{s_{NN}}$~=~200~GeV \dAu collisions 
\footnote{It would also be interesting to determine if the original Cronin data~\cite{Cronin:1974zm}
can be described
with a blast-wave fit to see whether hydrodynamics might account for the Cronin effect.}
and
heavy mesons.  Results from RHIC have shown a significant enhancement of electrons from
the decay of heavy mesons in 0--20\% central \dA collisions~\cite{Adare:2012yxa}.  If the enhancement observed in
\pA collisions is from a radial flow boost, then $D$ and $B$ mesons will be more enhanced 
because of their larger mass.  Blast-wave fits were done to the \dA spectra in Ref.~\cite{Adare:2013esx}
and was found to give a good description of the spectra~\cite{Sickles:2013yna}.
The blast-wave fit parameters from the light hadron spectra were used to determine a 
blast-wave prediction for the $D$ and $B$ meson spectra.  These spectra were compared to theoretical calculations from
Fixed-Order-Next-to-Leading-Log (FONLL) calculations~\cite{Cacciari:2001td,Cacciari:2012ny,fonllwp}
of $D$ and $B$ spectra.
The resulting nuclear modification factors for $D$ and $B$ mesons are shown in Fig.~\ref{fig:bwrda} (left).
In order to compare with the RHIC data which are for electrons from charm and bottom hadron decays,
information about the decay kinematics from PYTHIA~\cite{Sjostrand:2007gs} is used.  The resulting
electron $R_{dAu}$ for 0--20\% central \dAu collisions is shown in Fig.~\ref{fig:bwrda} (center)
compared with the measurement from Ref.~\cite{Adare:2012yxa}.
Given the uncertainties both on the data and the blast-wave parameters, the data and the 
calculation are in reasonable agreement.  
The calculations can also be done using the ALICE blast-wave parameters and the FONLL calculation
at 5.02~TeV.
At $\sqrt{s_{NN}}$~=~5.02~TeV the harder initial mesons spectra
will lead to a smaller enhancement.  The resulting $R_{pA}$ based on high multiplicity \pPb 
blast-wave parameters from ALICE\cite{Abelev:2013haa} is shown in Fig.~\ref{fig:bwrda} (right).
The data from ALICE are thus far only for minimum bias \pPb collisions so a direct comparison of the
calculation to the data is not possible at this time.

\begin{figure}
\centering
\includegraphics[width=0.56\textwidth]{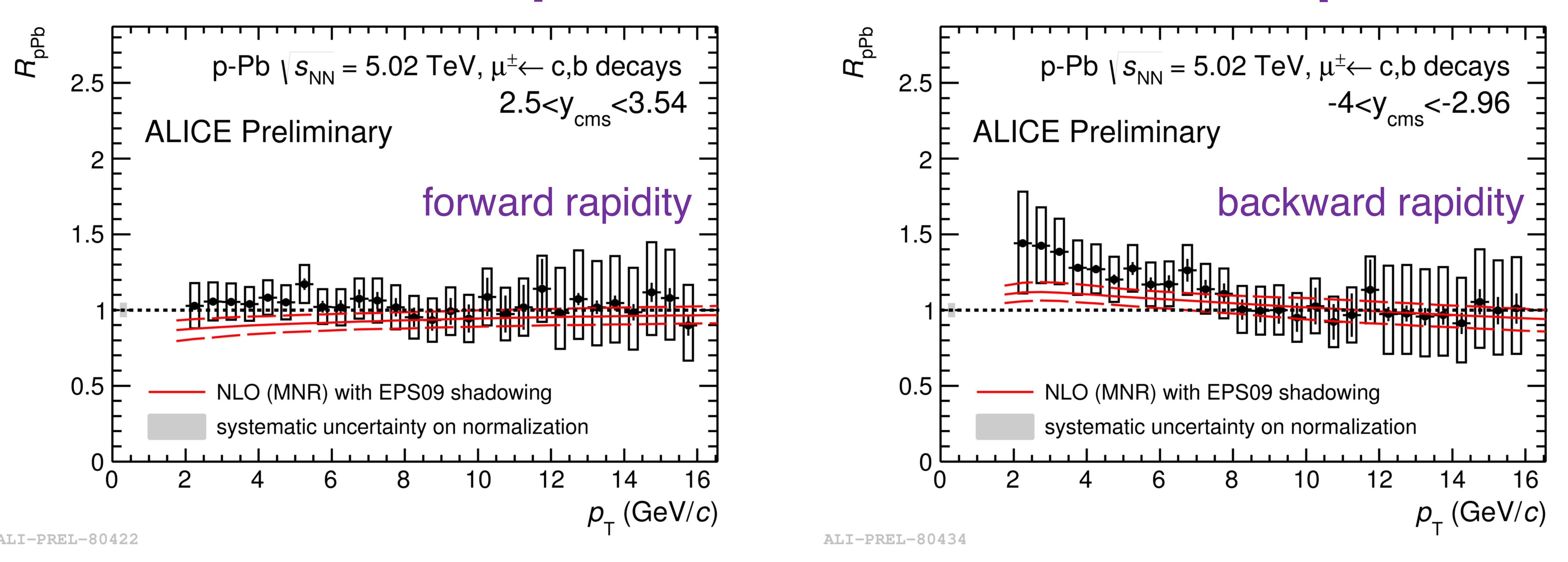}
\includegraphics[width=0.32\textwidth]{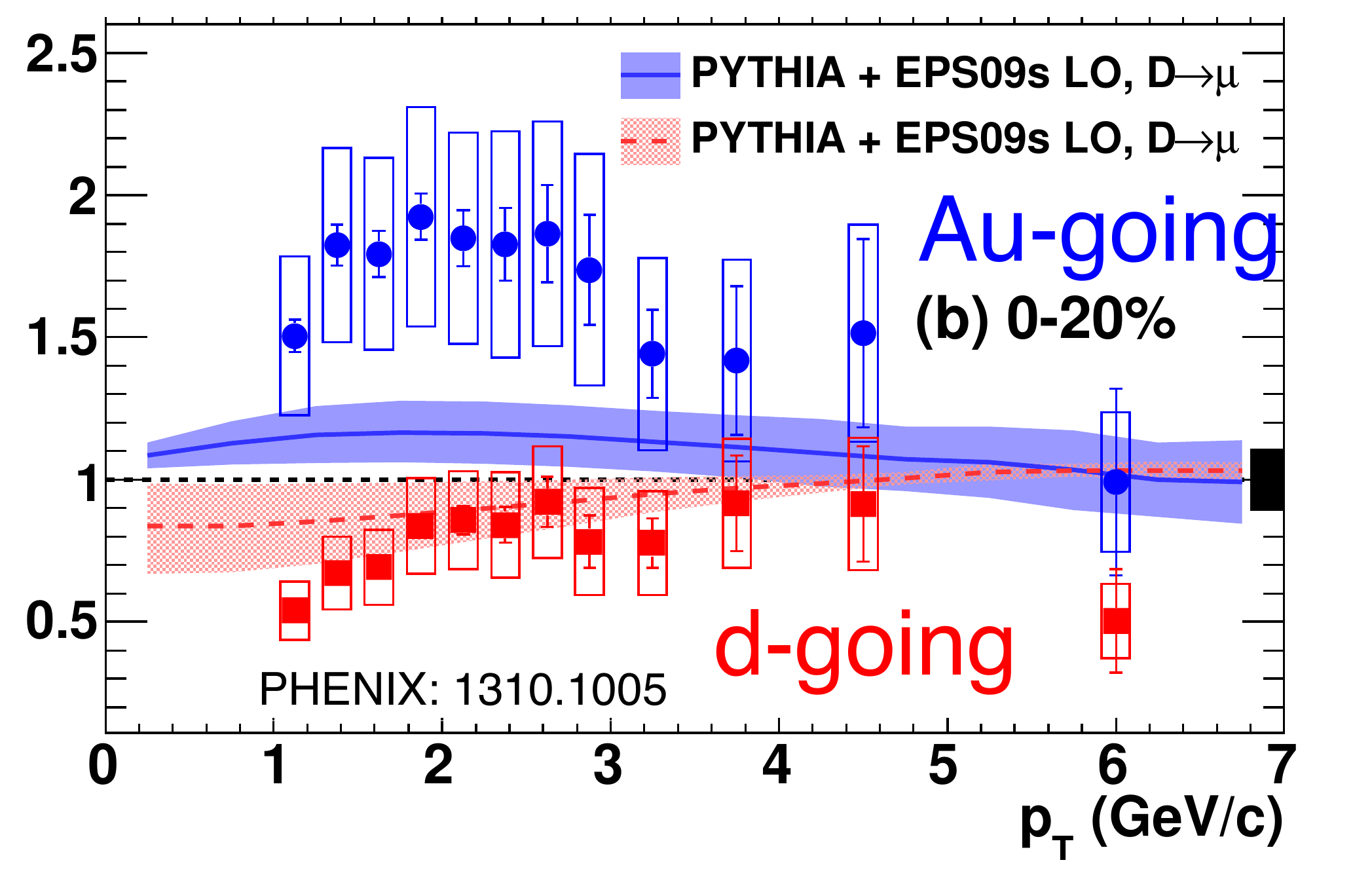}
\caption{Nuclear modification factors for $\mu^{\pm}$ from heavy meson decays.  Left and middle panels
show $R_{pPb}$ for forward (left) and backward (middle) in minimum bias \pPb collisions~\cite{Li:qm14}
and the right panel shows 
forward (squares) and backward (circles) $R_{dAu}$ in central \dAu collisions~\cite{Adare:2013lkk}.  
Curves on all panels show expectations
from EPS09~\cite{Eskola:2009uj}.
directions in \pPb collisions. and Ref.~\cite{Adare:2013lkk}.}
\label{fig:forwardhf}
\end{figure}

In asymmetric nuclear collisions, the particle production is also asymmetric around mid-rapidity.  In \dAu collisions
at $\sqrt{s_{NN}}$~=~200~GeV the maximum in $dN_{ch}/d\eta$ is near $\eta\approx$~2--3 in the Au-going
direction~\cite{Back:2004mr}.  Therefore, effects 
which are sensitive to the bulk particle production would  be expected to also
be strong in this region as well.  Correspondingly, in the 
in the $d$-going direction where there is less particle production, these effects would be expected to be reduced.
ALICE~\cite{Li:qm14} and PHENIX~\cite{Adare:2013lkk} have measured muons from the decay of 
heavy mesons on both sides of midrapidity, in 
addition to the midrapidity electron measurements discussed above.  The results are shown in 
Fig.~\ref{fig:forwardhf}.  Both experiments
observe an enhancement in the nucleus-going direction (denoted as ``backward"), 
which is perhaps somewhat larger in PHENIX than
in ALICE though the systematic uncertainties are large on both measurements.  
In the proton-going direction (denoted as ``forward") ALICE observes an $R_{pPb}$
consistent with unity while PHENIX observes a small suppression in the $d$-going direction.  The curves in 
Fig.~\ref{fig:forwardhf} show the expectations from nuclear PDFs~\cite{Eskola:2009uj,Helenius:2012wd}.
The enhancement in the Au-going direction at RHIC is larger than is expected from EPS09s and
the $d$-going $R_{dAu}$ is consistent with EPS09s expectations.  In ALICE,
the enhancement in the Pb-going direction is perhaps slightly larger than the expectation from EPS09, 
though the uncertainties on the data are large.

\pAu running at $\sqrt{s_{NN}}$~=~200~GeV is expected in 2015.  Since the \dAu data taking
in 2008  silicon vertex detectors have been installed
in both STAR and PHENIX.  This will enable the separation of decay electrons from charm and bottom which
will be sensitive to whether there is the mass dependence characteristic of radial flow as well as
better comparisons to nuclear parton distribution functions and LHC data.
Direct reconstruction of $D$ mesons has been done at ALICE~\cite{Li:qm14}.
$D$ mesons have been measured at RHIC with the 2003 \dAu data~\cite{Adams:2004fc}  and higher statistics 
measurements would be of great interest at RHIC both in \pAu and \dAu.

\section{Azimuthal Correlations in \pA}

The interest in azimuthal correlations in \pA collisions began with the observation of 
a long range correlation in pseudorapidity of particles close together in azimuth,
{\it the ridge}, in \pPb
collisions~\cite{CMS:2012qk}.  
The ridge has been understood in heavy-ion collisions as being due to azimuthal correlations
from the initial geometry of the collision followed by hydrodynamic evolution.
The discovery of a ridge in \pA collisions was immediately suggestive of collective 
behavior.
The discovery that there was also an away-side ridge under the recoiling
jet peak~\cite{Abelev:2012ola,Aad:2012gla} and that the correlations were larger in \dAu 
collisions than in \pPb collisions~\cite{Adare:2013piz} strongly suggested that the correlations were caused by
hydrodynamic flow as is the case in nucleus-nucleus collisions.  Another possible explanation
is that the back-to-back correlations  are due to Glasma diagrams in the Color-Glass Condensate~\cite{Dusling:2013oia}.
Over the last year much experimental activity has gone into trying to understand the origin of these
correlations.  Here we highlight some new results and discuss future opportunities.

The first results of $v_2$ in \pA systems were based on two particle azimuthal correlations.  This method has 
several limitations.  The main limitation is the influence of dijet
correlations on the $v_2$ measurement.  
Additionally, this method is insensitive to the number
of particles which are correlated.  Here we discuss how more sophisticated techniques are being
used to make qualitative improvements in the $v_N$ measurements in \pA collisions.

The most straightforward method removing dijet correlations is the {\it peripheral subtraction method}.  
In order to remove
dijet correlations from high multiplicity, or central events, the correlations are measured in 
peripheral events (low multiplicity events).
The peripheral subtraction technique relies on the assumption that modifications to jet correlations are
small compared to the correlations attributed to $v_N$.  On the near side, sensitivity to this effect can be
reduced by increasing the $\eta$ separation between the particles.  On the away side the jets are not well correlated
in $\eta$, so regardless of the $\eta$ separation between the particles the away side jet contribution
to two-particle correlations is approximately constant.
The measurement is then sensitive to differences in the dijet correlations between
central and peripheral events.  

This limitation is particularly applicable to the first measurements of $v_2$ in \dAu collisions~\cite{Adare:2013piz}.
There the required $\eta$ separation was between  0.5--0.7, limited by the size of the PHENIX charged particle 
acceptance.  
This separation is not enough to completely exclude the near side
jet.
In contrast, the LHC experiments and STAR have a wider midrapidity charged particle
 acceptance in $\eta$ allowing more complete suppression
of the near side jet correlations.  

\begin{figure}
\centering
\includegraphics[width=0.32\textwidth]{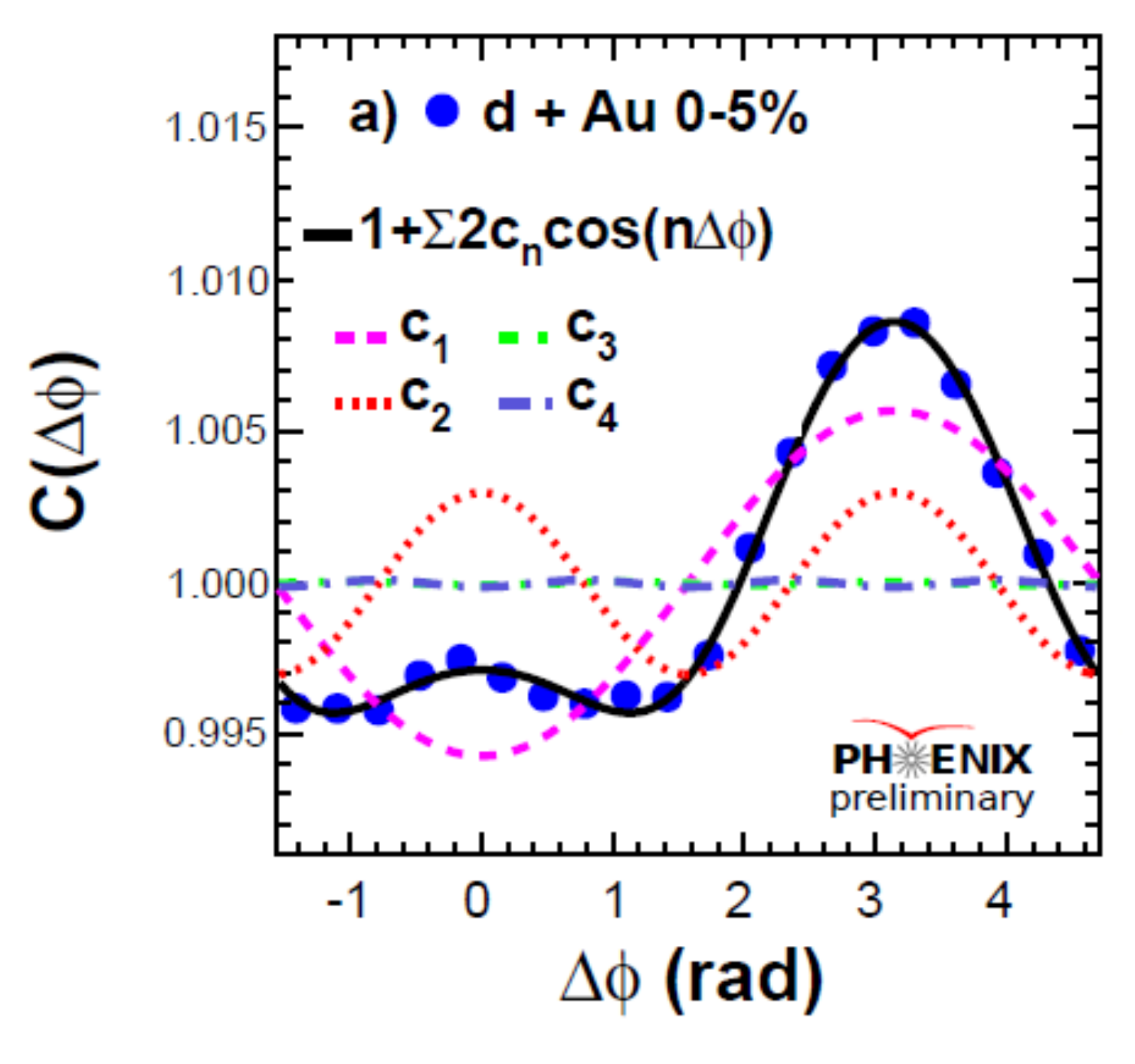}
\includegraphics[width=0.51\textwidth]{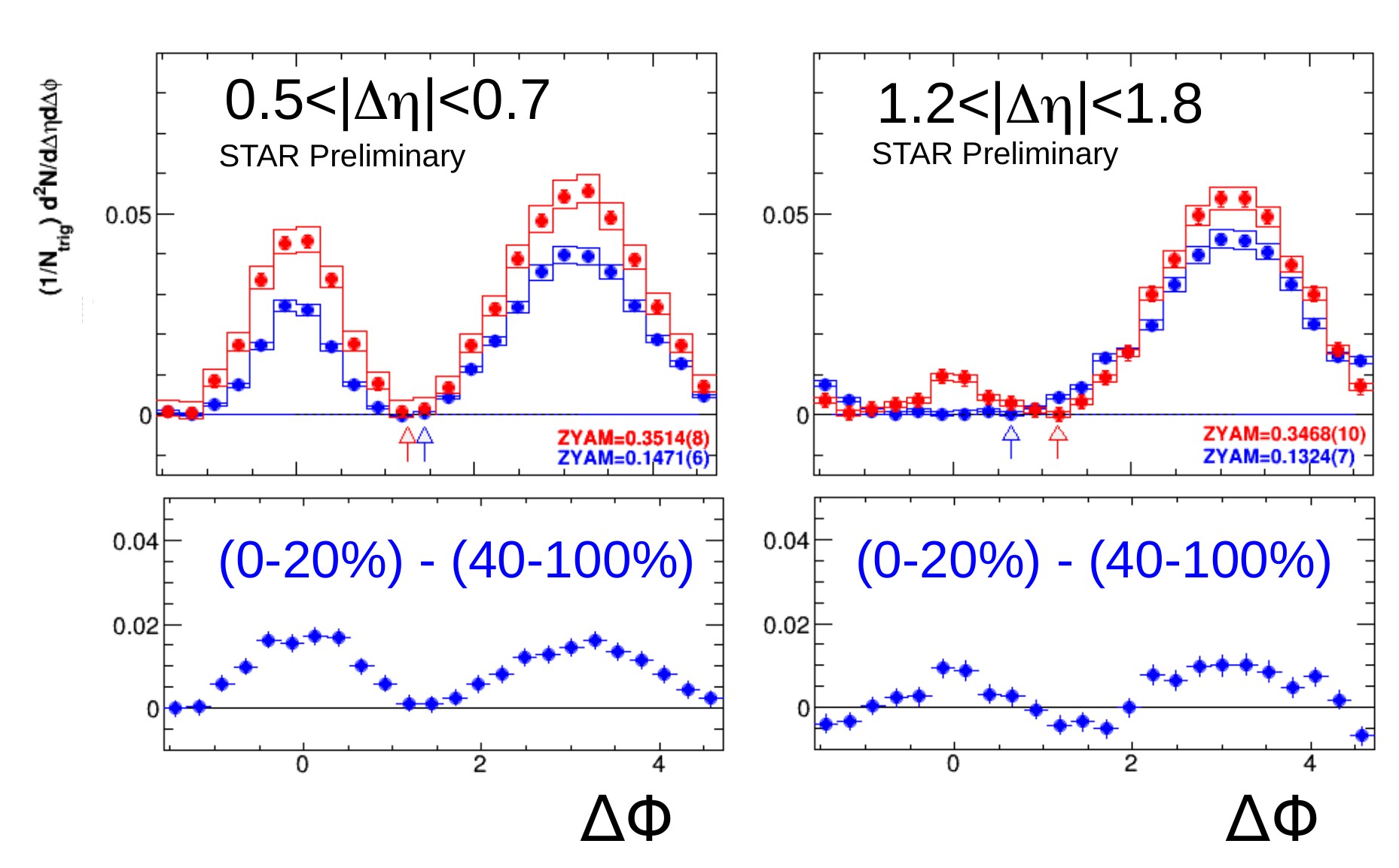}
\caption{
(left panel) Correlations of the $E_{T}$ measured by the d-going and Au-going MPC detectors 0--20\% central
\dAu collisions.  A ridge
correlation is seen for $|\Delta\eta|>$~6.
(top row, right two panels) 
Azimuthal correlation for 0-20\% central events (light points) and 40-100\% central events (dark points)
for two different $\Delta\eta$ selections: 0.5--0.7 (middle) and 1.2--1.8 (right).  The
background has been subtracted using the zero yield at minimum technique~\cite{Ajitanand:2005jj}.
(bottom row, right two panels) 
Results of a perpherial subtraction of the corresponding upper panels.  Excess at $\Delta\phi\approx 0$
and $\Delta\phi\approx\pi$ are seen in all $\Delta\eta$ ranges (there is a downward shift in the right panel).
Plot is from Ref.~\cite{Yi:qm14}.  It should be noted that the centrality selections used here are not the same 
was used in Ref.~\cite{Adare:2013piz}.}
\label{fig:starsub}
\end{figure}

While at midrapidity PHENIX has a limited accessible $\Delta\eta$ range for charged particles, 
long range correlations have been shown between 
midrapidity charged hadrons and the energy in the Au-going muon piston calorimeter 
(MPC)~\cite{Adare:2014keg,Huang:qm14}. 
A ridge has also been observed in energy correlations between the Au-going and d-going MPCs with 
$|\Delta\eta|>$~6~\cite{Huang:qm14}, as seen in Fig~\ref{fig:starsub}.
PHENIX has used the Au-going energy to determine an event plane in \dAu collisions and measure the $v_2$ of
charged hadrons near midrapidity with respect to this plane~\cite{Adare:2014keg}.  The event plane $v_2$ results
are slightly lower than the two-particle correlation results in Ref.~\cite{Adare:2013piz}.

The STAR detector has the ability to measure over a wider range
in pseudorapidity separation.  Results for central and peripheral correlations for a single \pT selection and
two $\Delta\eta$ selections are shown in Fig.~\ref{fig:starsub}.  Following a peripheral subtraction
near and away side correlations are observed in correlations with $|\Delta\eta|$ up to 1.8.
In this $\Delta\eta$ range the near side jet is completely removed in 40--100\% peripheral collisions and
a small near side correlation is observed in 0--20\% central collisions.
In order to explore the possible effects of dijet  modifications between the jets in central and peripheral
collisions, STAR
introducted a scaling factor to match the near side jets in the small $\Delta\eta$ region.  Applying this scaling 
factor to different $\Delta\eta$ selection causes the away side correlation to be reduced.  
There was discussion that this finding was inconsistent with the 
observation of collective effects in \dAu collisions.  This conclusion is
premature.  A study of the sensibility of this rescaling procedure in more \pT bins is called for.  Additionally, 
the validity of the peripheral subtraction method with a wide central bin (0--20\%) and a rather central ``peripheral''
bin (40--100\%) is not clear and has not been demonstrated.  
Given the focus of the community on the topic of angular correlations in very small collisions, these
and related studies should be given a high priority.
The ATLAS collaboration has employed a
similar technique~\cite{Radhakrishnan:qm14}, 
but both ridges remain.  A fuller understanding of this procedure
requires a systematic study as a function of \pT, centrality, and $\eta$ acceptance (the ATLAS $\eta$ acceptance 
is much wider than that of the STAR detector).

\begin{figure}
\centering
\includegraphics[width=0.65\textwidth]{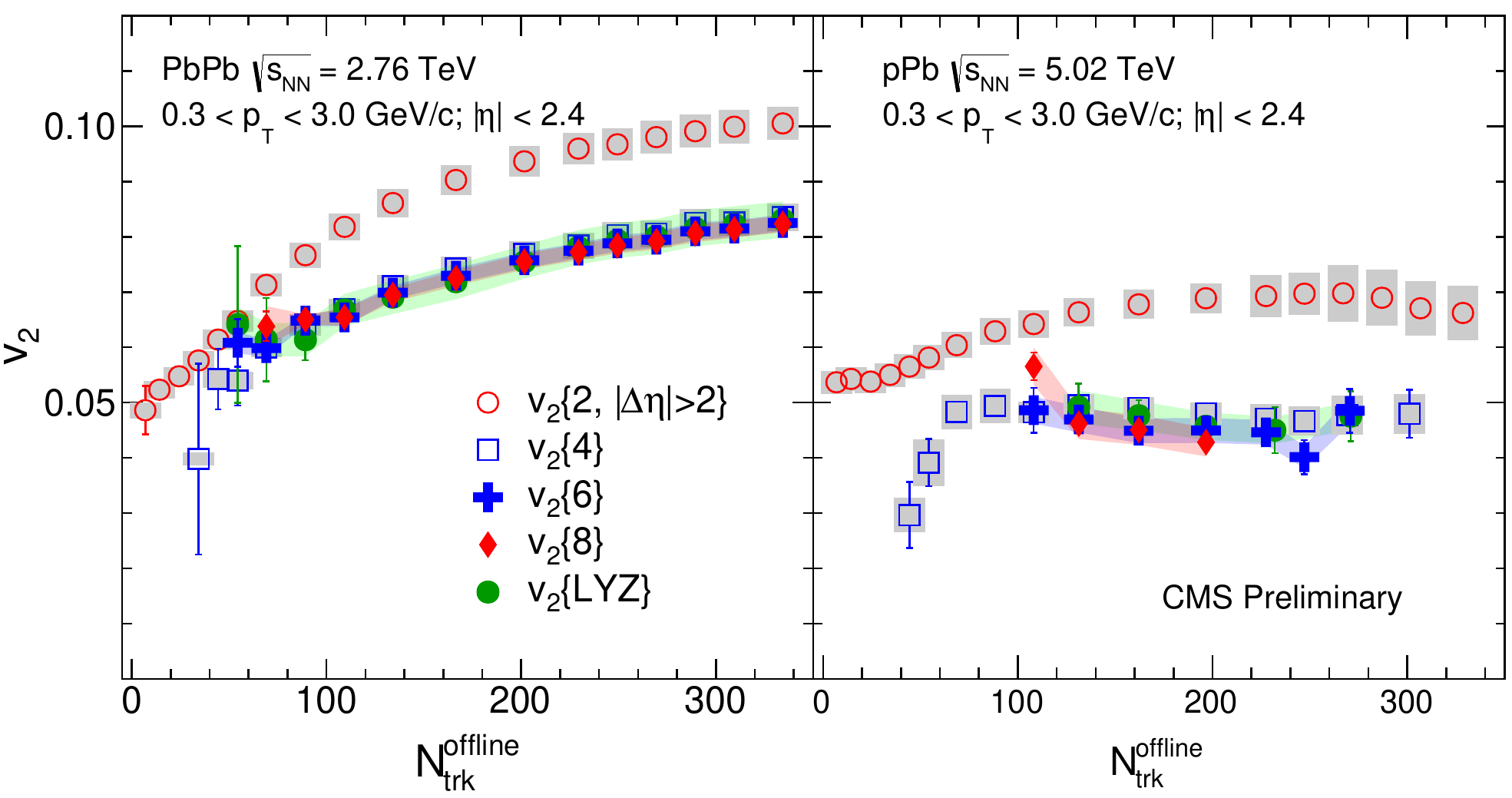}
\caption{$v_2$ in \PbPb (left) and \pPb (right) collisions as measured with 2, 4, 6 and 8 particle cumulants
as well as the Lee-Yang Zeros method~\cite{Wang:qm14}.  The $v_2$ values are plotted as a function of the 
number of offline tracks as measured by CMS.}
\label{fig:cmscumulants}
\end{figure}

Two-particle correlations alone are not sensitive to whether many or a few particles (for example
as is the case with jets) are correlated.  
In the Color-Glass Condensate model only a few particles in the event should  be correlated
by the Glasma diagrams, whereas if
hydrodynamics is the source of observed correlations all of the particles should be correlated.  
The event plane results discussed above suggest that many particles
are correlated rather than just a few.
At the LHC, CMS has investigated this by measuring $v_2$ with cumulants.  Results for \PbPb and \pPb are shown in 
Fig.~\ref{fig:cmscumulants}~\cite{Wang:qm14}.
The $v_2$ from two particle correlations with a rapidity separation between the two particles is found to be larger
than the $v_2\{4\}$ ($v_2$ as measured via four particle cumulants).  However increasing the order of the cumulants
from 4 to 8 does not change the observed $v_2$ in either \PbPb or \pPb.  Additionally, CMS has 
extracted $v_2$ from  the Lee-Yang
Zeros method~\cite{Bhalerao:2003xf} 
which involves correlations among all the particles in the event.  The $v_2$ extracted
via that method is also found to be consistent with the cumulant $v_2$ for four or more particles. These results are consistent
with the origin of the correlations in hydrodynamic phenomena and is not the behavior that is expected from
the Color-Glass Condensate where only a few particles would be expected to be correlated.

\begin{figure}
\centering
\includegraphics[width=0.55\textwidth]{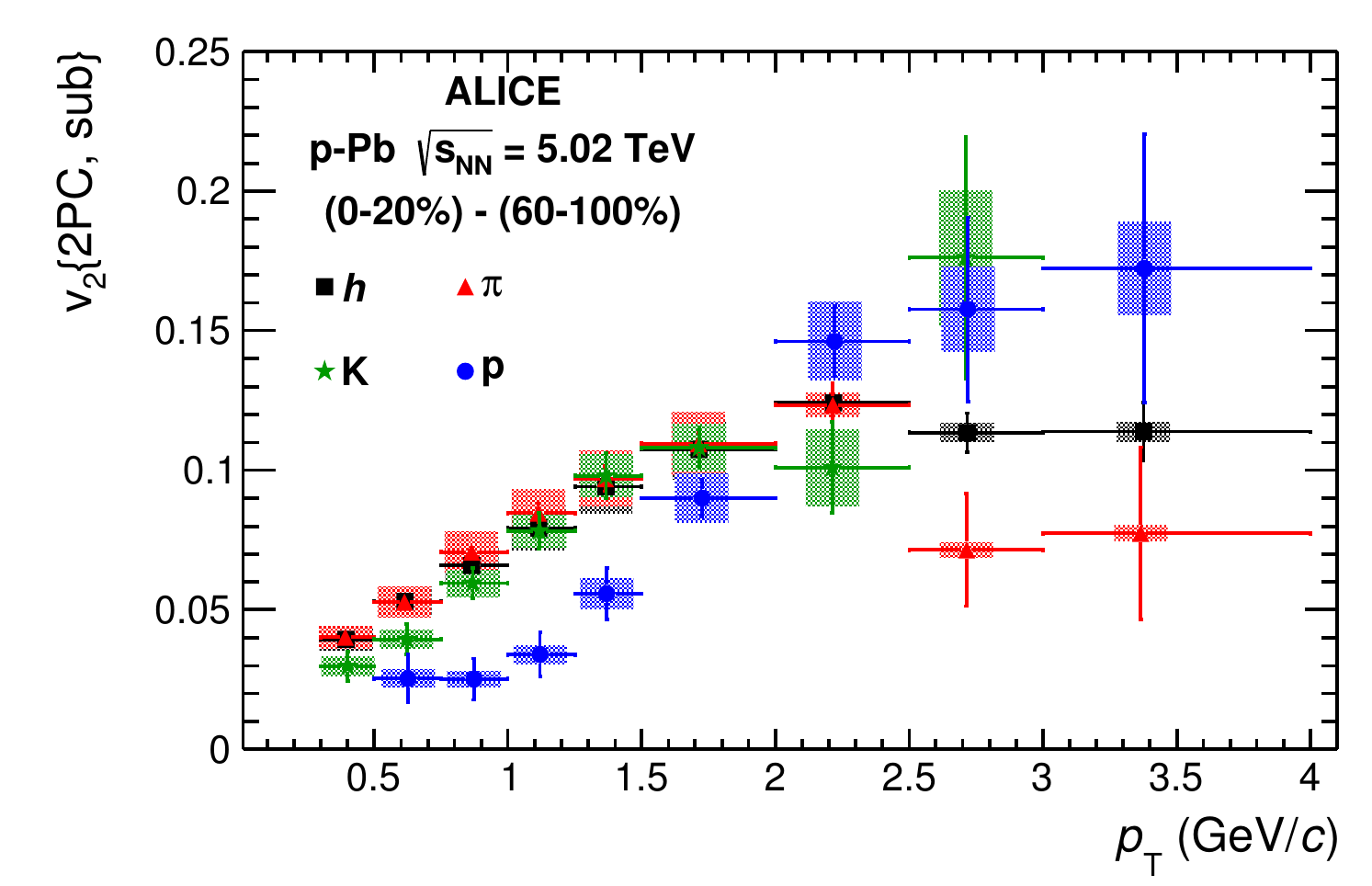}
\includegraphics[width=0.36\textwidth]{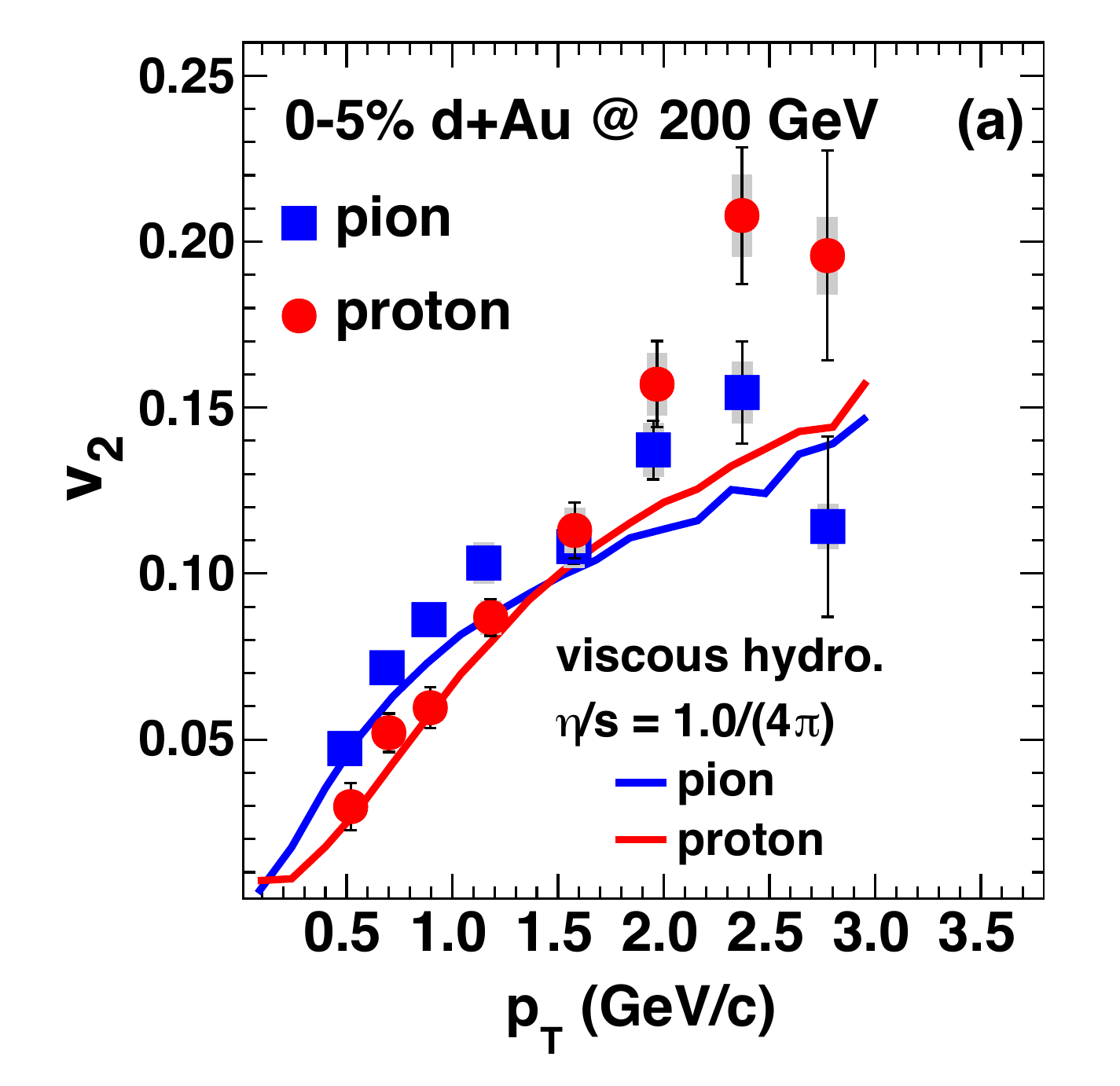}
\caption{$v_N$ for $\pi^{\pm}$ and $p,\bar{p}$ for high multiplicity \pPb collisions from ALICE~\cite{ABELEV:2013wsa} (left)  
and central \dAu collisions from PHENIX~\cite{Adare:2014keg} (right).
Shown on the PHENIX data is a hydrodynamic calculation from Ref.~\cite{Nagle:2013lja}.}
\label{fig:pidvn}
\end{figure}

Radial flow, in addition to modifications of the particle spectra discussed in Section~\ref{sec:spectra},
 should also affect the $v_N$ of hadrons in a mass dependent manner.  Measurements of $v_2$
for pions and (anti-)protons are now available in both \pPb (using the
subtraction method)~\cite{ABELEV:2013wsa} and \dAu collisions (using
the event plane method)~\cite{Adare:2014keg} and are shown
in Fig.~\ref{fig:pidvn}.  In both collision systems, the mass ordering of $v_2$ characteristic of
radial flow is observed.  The splitting is larger in \pPb collisions, consistent with 
larger radial flow in \pPb collisions.  A hydrodynamic calculation~\cite{Nagle:2013lja} is able to qualitatively
describe both the \pPb and \dAu results. However, the magnitude of the observed splitting in \pPb collisions is 
larger than that expected within the calculation.

If the correlations observed in \pA systems are a response to the initial collision geometry, then
changing the projectile nucleus should be able to change the measured $v_N$.  This was seen in
\dAu collisions where the $v_2$ in central \dAu collisions was found to be larger than in in \pPb
collisions~\cite{Adare:2013piz} possibly due to the initial elongated shape of the deuteron.
Pursuing that idea further, it might be possible to observe a large $v_3$ by colliding a triangluar 
nucleus with a large nucleus.  In the time since this conference, RHIC has successfully delivered $^3$He+Au
collisions at $\sqrt{s_{NN}}$~=~200~GeV.  Hydrodynamic calculations show some increase in the
expected $v_3$ of these collisions compared to \dAu (in \dAu no significant $v_3$ has been 
observed~\cite{Adare:2013piz,Adare:2014keg}), especially in very high multiplicity events where the
$^3$He nucleus could preferentally be oriented to maximize the triangularity of the collision 
region~\cite{Nagle:2013lja}.
As a further comparison, \pAu running at RHIC is expected in 2015.  The full collection of these three
small nuclei collided with Au will provide the most constraining tests as to the correlations
between the geometry of the initial state and the final state correlations.

\section{Conclusions}

There has been great interest in determining the role that hot nuclear matter effects play in 
\pA collisions.  A wealth of new measurements were  presented at this conference  aimed at
precisely this question.  The available data strongly suggests that the angular correlations observed
in these small collision systems involve many (or all) of the particles in the event, persist to large
pseudorapidity separations and are related to the collision geometry.  The results from single identified particle
measurements support this interpretation.  Mass dependent enhancement in the $R_{pA}$ measurements has been observed
and there are hints that this may extend to the heavy mesons.  The modifications to the particle composition 
does not seem to originate in jets, but rather in non-jet particles.  

The similarity of all these measurements to what has been seen in heavy-ion collisions is striking and suggests 
a common origin of phenomena in both systems.  There was much discussion at the conference on whether hydrodynamic
models could explain the totality of the data small and large system data.  First attempts at more
comprehensive calculations are being made~\cite{Schenke:2014zha} and more work is clearly necessary.
The ability to quantitatively extract $\eta/s$ from the
central heavy-ion data will be unconvincing without a clear understanding of the applicability of
hydrodynamics in small collision systems.
Since smaller collision systems can create less matter and will live for a shorter period of time they
potentially provide a good place to descriminate between models of the collision at very early times.
Upcoming $^3$He+Au and \pAu data to be taken at RHIC will be key to experimentally constraining the
physics of small collision systems.

While much of the interest in \pA has focussed on low-\pT physics, high \pT physics of hadrons, jets,
other hard probes and the centrality dependence of their production has also been of great interest.  
Effects are seen which are larger than expected from nuclear modifications to the parton distribution 
functions, the origin of which is not understood.  
Some of these effects are seen in the centrality dependent results and further investigation of the
centrality determination in \pA and the correlations between that determination and the measured
hard probes are under investigation.  The comparison between \pPb collisions at the LHC 
and the upcoming \pAu collsions at RHIC will be of great interest.
In contrast to heavy-ion collisions, no  evidence for
jet quenching in \pA has been observed.

\pA systems have proven to be much more interesting and surprising than originally envisioned 
and might be providing the most discriminating opportunity to push our understanding of
collective behavior in
heavy-ion physics to its limits.  Other surprising effects remain unexplained.  Upcoming \pA and $^3$He+Au
running at RHIC as well as more \pPb measurements will provide more insight into the physics of asymmetric nuclear 
collisions.

\section{Acknowledgements}
I thank the organizers for providing such an enjoyable and stimulating conference.
I would also like to thank all those who discussed their work with me in preparation for this talk.  I would 
like to thank the ALICE, ATLAS, CMS, PHENIX and STAR collaborations for providing me with the material 
necessary to put this talk together in a timely manner.
This work was supported by the DOE under contract number: DE-AC02-98CH10886.




\bibliographystyle{elsarticle-num}
\bibliography{sickles_qm14proceedings}







\end{document}